\documentclass[preprint, NumberedRefs]{JASA}
\usepackage{booktabs}
\usepackage{amsmath,mathtools}
\usepackage{amssymb}
\usepackage{amsthm}
\usepackage{mathrsfs}
\usepackage{physics}
\usepackage{bm, upgreek}
\usepackage{color}
\usepackage{tipa}

\usepackage{xcolor}
\definecolor{BrewerSetRed}{RGB}{228,26,28}
\definecolor{BrewerSetBlue}{RGB}{55,126,184}
\definecolor{BrewerSetGreen}{RGB}{77,175,74}
\definecolor{BrewerSetPurple}{RGB}{152,78,163}
\definecolor{BrewerSetOrange}{RGB}{255,127,0}
\definecolor{BrewerSetYellow}{RGB}{236,176,23}
\definecolor{BrewerSetBrown}{RGB}{166,86,40}
\definecolor{BrewerSetPink}{RGB}{247,129,191}
\definecolor{BrewerSetGray}{RGB}{153,153,153}

\usepackage{tikz}
\usetikzlibrary{shapes.geometric,positioning}
\DeclareRobustCommand\SolidLine[1]{
    \tikz[baseline=-0.6ex]\draw[very thick, #1] (0,0) -- (2em,0);
}
\DeclareRobustCommand\DashdottedLine[2]{
    \tikz[baseline=-0.6ex]\draw[very thick, dash pattern = on 5pt off 1.5pt on 3pt off 1.5pt on 5pt off 1.5pt on 3pt off 1.5pt on 5pt, #1] (0,0) -- (#2,0);
}

\usepackage{hyperref}
\usepackage{bookmark}

 \usepackage{tgtermes}
\begin{document}
 
\title[]{A k-space approach to modeling multi-channel parametric array loudspeaker systems}

\author{Tao Zhuang}
\affiliation{Key Laboratory of Modern Acoustics and Institute of Acoustics, Nanjing University, Nanjing 210093, China}

\author{Longbiao He}
\affiliation{Division of Mechanics and Acoustics, National Institute of Metrology, Beijing 100029, China}

\author{Feng Niu}
\affiliation{Division of Mechanics and Acoustics, National Institute of Metrology, Beijing 100029, China}

\author{Jia-Xin Zhong}
\email{Jiaxin.Zhong@psu.edu}
\affiliation{Graduate Program in Acoustics, College of Engineering, The Pennsylvania State University, University Park, PA 16802}
  


\author{Jing Lu}
\affiliation{Key Laboratory of Modern Acoustics and Institute of Acoustics, Nanjing University, Nanjing 210093, China}

 
 
\date{\today} 
 
\newpage
\begin{abstract}
	 \textbf{ABSTRACT}


\noindent Multi-channel parametric array loudspeaker (MCPAL) systems offer enhanced flexibility and promise for generating highly directional audio beams in real-world applications.
However, efficient and accurate prediction of their generated sound fields remains a major challenge due to the complex nonlinear behavior and multi-channel signal processing involved.
To overcome this obstacle, we propose a $k$-space approach for modeling arbitrary MCPAL systems arranged on a baffled planar surface. 
In our method, the linear ultrasound field is first solved using the angular spectrum approach, and the quasilinear audio sound field is subsequently computed efficiently in $k$-space. 
By leveraging three-dimensional fast Fourier transforms, our approach not only achieves high computational and memory efficiency but also maintains accuracy without relying on the paraxial approximation. 
For typical configurations studied, the proposed method demonstrates a speed-up of more than four orders of magnitude compared to the direct integration method.
Our proposed approach paved the way for simulating and designing advanced MCPAL systems.

\end{abstract}

\maketitle



\section{Introduction}
\label{sec:intro}

Parametric array loudspeakers (PALs) generate highly directional audio beams through nonlinear interactions of intense ultrasonic waves.\cite{Westervelt1963, Zhong2024} 
When a PAL radiates two harmonic waves, a phased array PAL composed of multiple PALs has been widely utilized in sound field control applications, such as beam steering and focusing.\cite{Shi2014, Gan2006, Ogami2019, Zhong2022a, Tanaka2010, Li2021}
Compared to phased arrays that only control the phase of each element, more advanced beamforming can be achieved through intricate multi-channel processing. 
This approach enables independent control of both phase and amplitude for each element, allowing for more flexible beam control. 
When each element in such a system consists of a micro-machined PAL unit,\cite{Lee2009, Je2015} the system can be referred to as a multi-channel PAL (MCPAL) system.\cite{Guasch2018, Zhu2025} 
In comparison to phased PAL arrays, MCPAL systems offer broader application potential, including sound field zoning control and the realization of length-limited PAL implementations.\cite{Geng2022, Geng2023, Zhuang2024, Zhu2025, Nomura2012, Nomura2025}
Accurate simulation of audio sound field radiation from MCPAL systems is essential for efficient design optimization.
However, such modeling remains highly challenging due to the intrinsic nonlinearities and structural complexity of these systems.

The sound field generated by a MCPAL system can be categorized into three distinct regions: the near field, the Westervelt far field, and the inverse-law far field.\cite{Zhong2021} 
Recent studies have focused on accurately modeling the audio sound generated by a MCPAL system in the inverse-law far field, such as the convolution directivity model.\cite{Shi2013, Shi2014, Shi2015, Gan2006, Guasch2018, Zhong2023, Mei2025} 
However, the audio sound in the near field and the Westervelt far field is more complex compared to the inverse-law far field, presenting a significant challenge for modeling these regions.\cite{Zhong2021} 
In the Westervelt far field, the acoustic waves generated by the MCPAL system can be accurately modeled using the Westervelt equation. 
In the near field, the audio sound generated by the MCPAL system exhibits increased complexity; however, it can be derived through algebraic corrections to the quasilinear solution of the Westervelt equation. \cite{Cervenka2022}
Consequently, for audio sound generated by the MCPAL system, the main computational task involves calculating the quasilinear solution of Westervelt equation.

Since the ultrasound level produced by a PAL is generally restricted within a safety range,\cite{Gan2012} a quasilinear approximation can be assumed to simplify the modeling based on Westervelt equation.\cite{Zhong2020} 
Under this framework, the calculation of the audio sound field requires numerical evaluations of five-fold integrals, which we refer to as the direct integration method (DIM) in this study.\cite{Zhong2024} 
Gaussian beam expansion (GBE) methods have been proposed to simplify the calculation, but they are inaccurate even in the paraxial region, especially at low audio frequencies, because the main lobe of the virtual sources spreads outside the paraxial region.\cite{Ye2010, Cervenka2013, Zhuang2023, Zhu2023} 
To address this, a method known as the spherical wave expansion (SWE) was proposed.\cite{Zhong2020, Zhong2021} 
By representing the Green's function as a superposition of spherical harmonics, the computation can be simplified by leveraging azimuthal symmetry. 
Moreover, an improved SWE method using Zernike polynomials can model a steerable PAL without azimuthal symmetry.\cite{Zhong2022}
However, the SWE method using Zernike polynomials is inherently limited to circular apertures and becomes inadequate for modeling non-circular overall geometries, such as rectangles. 
In practice, MCPAL systems often approximate such shapes by arranging many small circular emitters.
This discrete assembly introduces irregular gaps between elements, resulting in substantial variations in the effective radiation surface. 
Such deviations from an idealized continuous aperture pose serious challenges for accurately predicting the audio sound fields generated by these MCPAL systems.\cite{Gan2006, Shi2011, Shi2015, Zhong2023b}

The audio sound generated by conventional multi-channel electrodynamic loudspeaker systems is typically expressed as the linear superposition of the waves generated by individual loudspeakers.\cite{Choi2002} 
However, modeling MCPAL systems consisting of multiple elements presents two unique challenges. 
Firstly, the element of the MCPAL cannot be modeled as a point source, as the ultrasound wavelength is typically comparable to or smaller than the size of the element used in real-world applications. 
Secondly, the audio sound is generated through nonlinear processes. 
Consequently, the output of the MCPAL is not merely the sum of the audio sounds produced by each individual element; it also includes components arising from the nonlinear interactions between two ultrasound waves at different frequencies emitted by distinct elements.\cite{Zhuang2024, Zhu2025, Zhuang2025}
A recent investigation addressed this issue by decomposing MCPAL's acoustic radiation into the contributions of individual elements and the effects of coupling between these elements.\cite{Zhuang2025}
By combining the SWE method and the addition theorem, this method provides a way to simulate the audio sound radiated by a MCPAL system composed of circular elements. 
However, a significant drawback of this approach lies in its computational inefficiency when applied to systems comprising a large number of elements.

To enable the simulation of the audio sound radiated by a general MCPAL system with lower computational resources, this work presents a computationally efficient method  by combining the Angular Spectrum Approach (ASA) \cite{Zeng2008} and $k$-space method.\cite{Mast2001, Tabei2002}
Building on Westervelt equation's quasilinear solution,\cite{Gan2012} the ultrasound is calculated by the ASA, which is suitable for arbitrary radiation surface geometries set on a baffle. 
Then the frequency-domain $k$-space method is employed to transform the computationally intensive volume integral into a $k$-space multiplication via fast Fourier transform (FFT).
This approach eliminates geometric constraints while dramatically reducing computational costs compared to conventional DIM,\cite{Zhong2021} enabling efficient simulation of MCPAL systems. 
To demonstrate its effectiveness, numerical results are provided for a representative MCPAL system and the efficiency is analyzed.

\section{Theory}
\subsection{Problem formulation}
\label{sec:theory_problemFormulation}

\begin{figure}[htb]
\includegraphics[width = 0.5\textwidth]{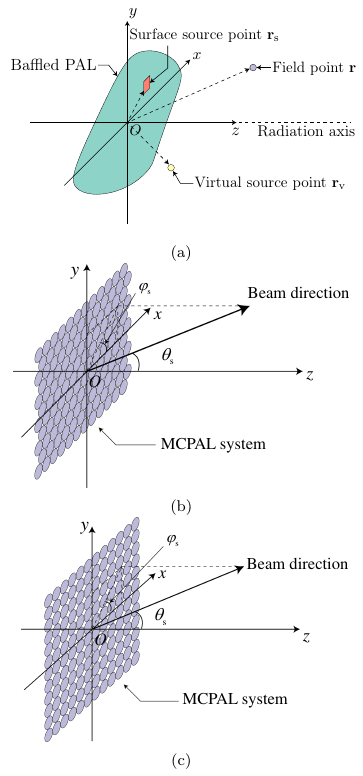}
\caption{\label{fig:sketch_PalArray}{
(a) The sketch of a PAL with an arbitrary radiation surface that is baffled in the plane $Oxy$.
(b--c) Sketch of (b) closely packed and (c) uniformly spaced MCPAL systems consisting of circular units.}}
\end{figure}

As shown in Fig.\,\ref{fig:sketch_PalArray}\,(a), a PAL is assumed to have an arbitrary radiation surface that is baffled in the plane $Oxy$. 
Under this configuration, MCPAL systems [Figs.~\ref{fig:sketch_PalArray}(b--c)] composed of multiple circular elements can be modeled by assuming that each element vibrates uniformly, while no vibration occurs outside these elements.
The coordinate systems $Oxyz$ is established with the $z$-axis is perpendicular to the radiation surface of the PAL. 
The PAL generates two harmonic ultrasound waves at frequencies $f_1$ and $f_2$ ($f_1<f_2$). 
the vibration velocity on the surface of the PAL is
\begin{equation}
    v_{\mathrm{I},z} \qty(\vb{r}_\mathrm{s})
    = v_{1,z} \qty(\vb{r}_\mathrm{s}) \mathrm{e}^{- \mathrm{i} \omega_1 t} 
    + v_{2,z} \qty(\vb{r}_\mathrm{s}) \mathrm{e}^{- \mathrm{i} \omega_2 t}, 
    \label{eq:PAL_velocity}
\end{equation}
where i is the imaginary unit, $\vb{r}_\mathrm{s} = (x_\mathrm{s}, y_\mathrm{s}, 0)$ is the surface source point, $v_i$ ($i = 1,2$) is the amplitude of the vibration velocity at $f_i$, and $\omega_i = 2 \pi f_i$ is the angular frequency of the ultrasound. 

The ultrasound pressure at frequency $f_i$ generated by the PAL at a virtual source point $\vb{r}_\mathrm{v} = (x_\mathrm{v}, y_\mathrm{v}, z_\mathrm{v})$ is denoted by $p_i \qty(\vb{r}_\mathrm{v})$, which can be obtained using the Rayleigh integral, shown as 
\begin{equation}
    p_i \qty(\vb{r}_\mathrm{v}) 
    = -2 \mathrm{i} \rho_0 \omega_i 
    \iint_S v_i \qty(\vb{r}_\mathrm{s}) g_i \qty(\vb{r}_\mathrm{v}, \vb{r}_\mathrm{s})
    \dd^2\vb{r}_\mathrm{s}
    ,
    \label{eq:ultra_rayleigh}
\end{equation}
where $\rho_0$ is the air density and 
\begin{equation}
    g_i \qty(\vb{r}_\mathrm{v}, \vb{r}_\mathrm{s})
    = \frac{\exp \qty( \mathrm{i} k_i \abs{\vb{r}_\mathrm{v} - \vb{r}_\mathrm{s}} )}{4 \pi \abs{\vb{r}_\mathrm{v} - \vb{r}_\mathrm{s}}}
    \label{eq:Green's_func}
\end{equation}
is the Green's function. 
Here, $k_i = \omega_i / c_0 + \mathrm{i} \alpha_i$ is the wavenumber of the ultrasound at frequency $f_i$ where $c_0$ is the sound velocity is air and $\alpha_i$ is the sound attenuation coefficient at frequency $f_i$. 
According to the quasilinear solution,\cite{Zhong2021} 
the audio sound radiated by the PAL can be considered as a superposition of the pressure radiated by infinite virtual audio sources in air with the source density of
\begin{equation}
q (\vb{r}_\mathrm{v}) 
= - \frac{\mathrm{i} \beta \omega_\mathrm{a}} {\rho_0^2 c_0^4} 
    p_1^* \qty(\vb{r}_\mathrm{v})
    p_2 \qty(\vb{r}_\mathrm{v}), 
\label{eq:PAL_virtual_source}
\end{equation}
where $\beta$ is the nonlinearity coefficient, 
the superscript ``*'' denotes the complex conjugate, 
 $\omega_{\mathrm{a}} = 2 \pi f_\mathrm{a}$, and $f_\mathrm{a} = f_2-f_1$ is the audio frequency. 
Then the audio sound at a field point $\vb{r} = (x, y, z)$ can be obtained by substituting Eq.\,(\ref{eq:PAL_virtual_source}) into\cite{Zhong2020}
\begin{equation}
 p_\mathrm{a} \qty(\vb{r}) 
 = - \mathrm{i} \rho_0\omega_\mathrm{a} 
 \iiint_V{
     q \qty(\vb{r}_\mathrm{v}) g_\mathrm{a} \qty(\vb{r}, \vb{r}_\mathrm{v})
     \dd[3]{\vb{r}_\mathrm{v}}
 },
\label{eq:audio_volume}
\end{equation}
where $k_\mathrm{a} = \omega_\mathrm{a} / c_0$ is the wavenumber of the audio sound. 

In practical applications, multiple ultrasonic transducer units are often combined to form a MCPAL system. 
Then the acoustic field radiated by the MCPAL system can be manipulated flexibly by controlling the amplitude and phase of individual units.
In practical implementations, the ultrasonic transducers comprising an MCPAL system typically have diameters ranging from 1\,cm to 2\,cm.\cite{Shi2015, Zhu2023, Zhuang2025, Zhu2025} 
Considering that the wavelength of 40\,kHz ultrasound in air is approximately 8\,mm, these transducer dimensions are acoustically pertinent relative to the wavelength. 
As a result, the configuration of the arrangement has a considerable impact on the acoustic field produced by the MCPAL system. 
This investigation focuses on two primary configurations: the closely packed and the uniformly spaced arrangements, as depicted in Fig.\,\ref{fig:sketch_PalArray}\,(b)\,and\,(c), respectively.
Both far-field simulation results and experimental data have shown that the audio sound fields radiated by MCPAL system vary depending on the arrangement. \cite{Shi2015, Zhu2023}.

\subsection{Ultrasound field\label{sec:theo_ultrasound}}
According to the ASA,\cite{Zeng2008} the ultrasound pressure calculated by the Rayleigh integral shown as Eq.\,(\ref{eq:ultra_rayleigh}) is a convolution, where the source is located in the $Oxy$ plane such that
\begin{equation}
 p_i \qty(\vb{r}) 
 = -2 \mathrm{i} \rho_0 \omega_i v_i \qty(x, y, 0) \ast_x\ast_{y} g_i \qty(\vb{r}, \vb{0}), 
 \label{eq:ultra_conv}
\end{equation}
where ``$\ast_x$'' and ``$\ast_y$'' represent the  convolution operator in $x$ and $y$ directions, respectively. 
Applying a two-dimensional (2D) Fourier transform to both sides of Eq.\,(\ref{eq:ultra_conv}) transforms the formula into the spatial frequency domain, 
\begin{equation}
    P_i \qty( k_x, k_y, z) 
    = -2 \mathrm{i} \rho_0 \omega_i V_i \qty(k_x, k_y, 0) G_i \qty(k_x, k_y, z), 
 \label{eq:ultra_conv_spatial}
\end{equation}
where $V_i(k_x,k_y,0) \equiv \mathcal{F}_x\mathcal{F}_y[v_i(x,y,0)]$ and $G_i(k_x,k_y,z) \equiv \mathcal{F}_x\mathcal{F}_y[g_i(\vb{r}, \vb{0})]$, where $\mathcal{F}_x$ and $\mathcal{F}_y$ is the Fourier transform in the $x$\,direction and $y$\,direction, respectively. 
It is noted that $G_i(k_x,k_y,z)$ has a closed-form solution of\cite{Zeng2008}
\begin{equation}
    G_i \qty(k_x, k_y, z) 
    = \frac{\mathrm{i} \exp(\mathrm{i} \sqrt{\Re(k_i)^2 - k_x^2 - k_y^2} \abs{z})}{2 \sqrt{\Re(k_i)^2 - k_x^2 - k_y^2}},
\end{equation}
where the operator $\Re(\vdot)$ extracts the real part.
The Fourier transform decomposes the diffracted wave into the superposition of plane waves, 
where $(k_x, k_y)$ are the transverse wavenumbers. 
The product of $V_i (k_x, k_y, 0)$ and $G_i (k_x, k_y, z)$ describes the angular spectrum of the propagating wave at $z$ plane. 
The ultrasound pressure is then obtained from the inverse 2D Fourier transform of $P_i \qty(k_x, k_y, z)$. 
When considering the sound attenuation, $G_i \qty(k_x, k_y, z)$ is multiplied by an exponential term \cite{Zeng2008} 
\begin{equation}
    A_i \qty(k_x, k_y, z) 
    = \exp \qty[ - \frac{\alpha_i \Re(k_i) \abs{z}}{\sqrt{\Re(k_i)^2 - k_x^2 - k_y^2}} ]. 
    \label{eq:sound_atten}
\end{equation}


\subsection{Audio sound field}
\label{sec:theo_audio_sound}
The audio sound calculated by the volume integral of the DIM given by Eq.\,(\ref{eq:audio_volume}) is a three-dimensional (3D) convolution, which can be expressed as
\begin{equation}
    p_\mathrm{a} \qty(\vb{r}) 
    = - \mathrm{i} \rho_0 \omega_\mathrm{a} q \qty(\vb{r}) \ast_x\ast_y\ast_{z} g_\mathrm{a} \qty(\vb{r}, \vb{0}). 
    \label{eq:conv_audio_sound}
\end{equation}
Applying a 3D Fourier transform to both sides of Eq.\,(\ref{eq:conv_audio_sound}) transforms the formula into the $k$-space domain, 
\begin{equation}
    P_\mathrm{a} \qty(\vb{k}) 
    = - \mathrm{i} \rho_0\omega_\mathrm{a} 
 \mathcal{F}_x\mathcal{F}_y\mathcal{F}_z \qty[ \iiint_V{
     q \qty(\vb{r}_\mathrm{v}) g_\mathrm{a} \qty(\vb{r}, \vb{r}_\mathrm{v})
     \dd[3]{\vb{r}_\mathrm{v}}} ]
     = - \mathrm{i} \rho_0 \omega_\mathrm{a} Q \qty(\vb{k}) G_\mathrm{a} \qty(\vb{k}), 
\end{equation}
where $\vb{k} = (k_x, k_y, k_z)$ is a point in the $k$-space domination, $Q(\vb{k}) \equiv \mathcal{F}_x\mathcal{F}_y\mathcal{F}_z [q(\vb{r})]$, and $G_\mathrm{a} \qty(\vb{k}) \equiv  \mathcal{F}_x\mathcal{F}_y\mathcal{F}_z [g_\mathrm{a}(\vb{r}, \vb{0})]$. 
The Fourier transform decomposes the volume integral of the DIM in the real space ($\vb{r})$ into the multiplication of two functions in $k$-space ($\vb{k}$). 
The product of the virtual source spectrum $Q \qty(\vb{k})$ and the spectrum of the Green's function $G_\mathrm{a} (\vb{k})$ describes the spectrum of the audio sound pressure in $k$-space domain. 
The audio sound pressure is then obtained from the inverse 3D Fourier transform of $P_\mathrm{a} \qty(\vb{k})$.
It is worth noting that no paraxial approximation is introduced in the above analysis; therefore, the proposed method retains the full accuracy of the DIM.

\subsection{Computational techniques and complexity analysis}
\label{sec:complex_analy}

\begin{figure}[htb]
\includegraphics[width = 0.6\textwidth]{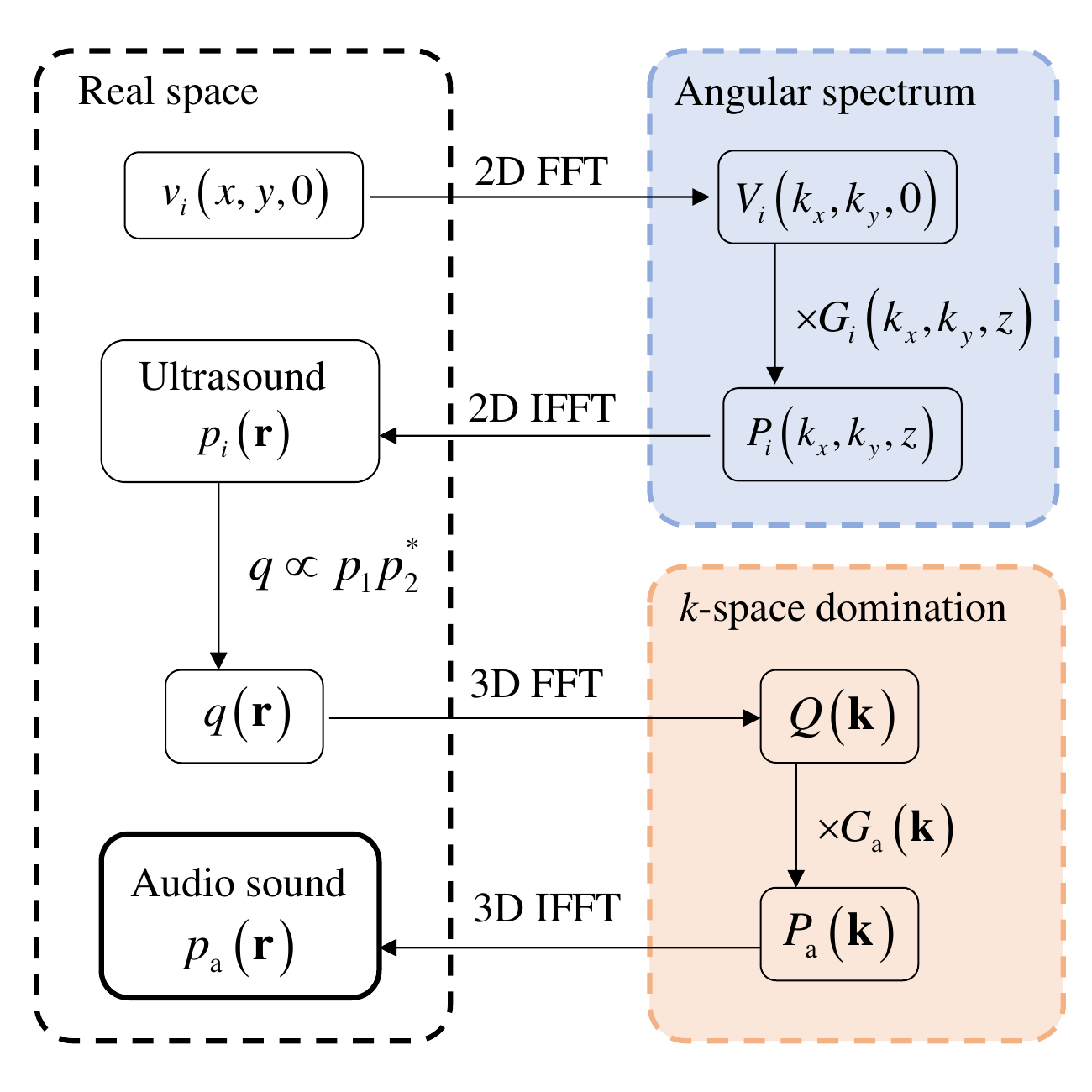}
\caption{\label{fig:diagram_kspace}{The diagram of modeling MCPAL systems using the proposed $k$-space method.}}
\end{figure}

To provide a clearer depiction of the proposed approach, Fig.\,\ref{fig:diagram_kspace} presents a detailed flowchart illustrating the proposed approach.
As discussed in Sec.\,\ref{sec:theory_problemFormulation}, computing the audio sound generated by an MCPAL system using the DIM in real space requires substituting the Rayleigh integral [Eq.\,(\ref{eq:ultra_rayleigh})] and the virtual volume source expression [Eq.\,(\ref{eq:PAL_virtual_source})] into the volume integral [Eq.\,(\ref{eq:audio_volume})]. 
As a result, the DIM requires the evaluation of five-fold integral in total.
However, the proposed method avoid the five-fold integral by using the Fourier transform, and the Fourier transform can be efficiently calculated using FFT.
Therefore, the proposed method using 3D FFT is much more computational efficient than the DIM. 

\begin{figure}[htb]
\includegraphics[width = 0.5\textwidth]{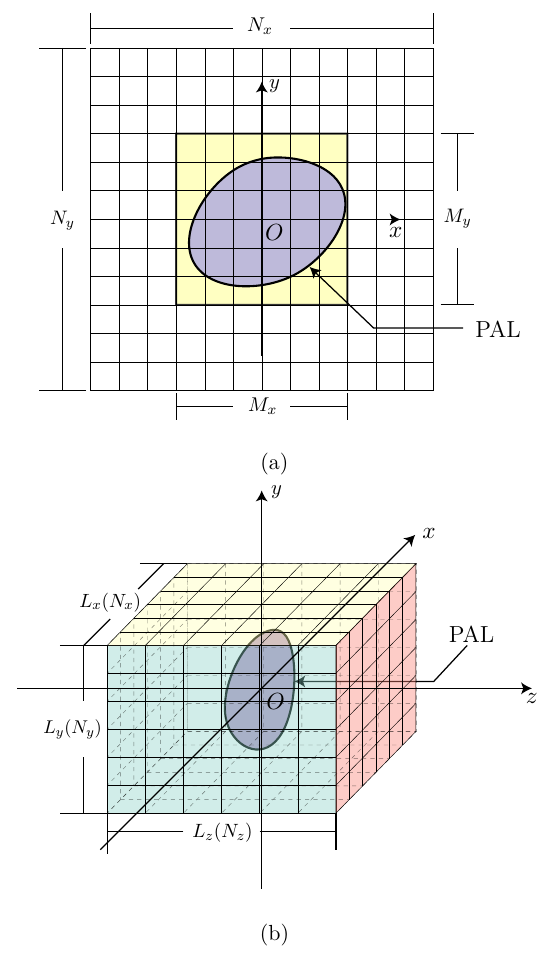}
\caption{\label{fig:numerical_grid_kspace}
{Schematic of mesh grids for numerically calculating (a) the (two-fold) Rayleigh integral for ultrasound and (b) the (three-fold) volume integral for audio sound. 
(a) The Rayleigh integral of DIM is computed over the yellow region (enclosing the purple radiation surface) using an $M_x M_y$ grid, while the surrounding ultrasound field employs ASA with a finer $N_x N_y$ grid.
(b) Volumetric integration is confined to an $L_x L_y L_z$ domain covering the ultrasound's main energy, discretized as $N_x N_y N_z$.}}
\end{figure}

Existing methods for simulating MCPAL systems often face limitations: for instance, the GBE method can only model systems composed of rectangular or elliptical elements, while the SWE method is restricted to circular-element configurations. 
To enable direct comparison of computational complexity across existing methods, this study focuses on an MCPAL system comprising $N_\mathrm{E}$ circular elements.
The computational efficiency of the proposed method is analyzed as follows. First, for the Rayleigh integral in Eq.\,(\ref{eq:ultra_rayleigh}), we define the source plane's numerical integration area as an $M_x M_y$ yellow grid region encompassing the PAL, as illustrated in Fig.\,\ref{fig:numerical_grid_kspace}\,(a).
It is noted that the Rayleigh integral requires an integration domain that fully covers the PAL surface, while the ASA necessitates an extended computational region to ensure sufficient accuracy. 
As illustrated by the outer boundary in Fig.\,\ref{fig:numerical_grid_kspace}\,(a), the ASA calculation domain employs an $N_x N_y$ mesh grid.
For the volume integral in Eq.\,(\ref{eq:audio_volume}), the infinite integration domain is truncated to a finite computational volume of dimensions $L_x L_y L_z$. 
This volume is discretized into $N_x N_y N_z$ grid points with spatial intervals $\Delta_x$, $\Delta_y$ and $\Delta_z$ along each axis, respectively [Fig.\,\ref{fig:numerical_grid_kspace}\,(b)].
It is worth noting that the proposed method is capable of obtaining the audio sound over a computational grid consisting of $N_x N_y N_z$ points, achieving a time complexity of $\mathcal{O}(N_x N_y N_z \log(N_x N_y N_z))$, which is dominated by the 3D FFT operations. 
In contrast, the computational complexity associated with utilizing the DIM to obtain audio sound at a single field point, which involves a five-fold integral, is expressed as $\mathcal{O}(M_x M_y N_x N_y N_z)$.
To obtain the full audio sound solution at all grid points, the complexity of the DIM increases to $\mathcal{O}(M_x M_y N_x^2 N_y^2 N_z^2)$, which is computationally intractable for dense discretization.

Table\,\ref{tab:DiffSimu_complex} summarizes the space and time complexity required for different methods to calculate the audio sound field distribution across the entire space ($N_x N_y N_z$ field points) and the regions where accurate calculations are possible. 
The memory demands of the output array, which stores audio signals at each of the $N_x N_y N_z$ field points, must be considered significant. 
Consequently, we assess performance by examining the total space complexity, which includes auxiliary space used during calculations, as well as the input and output space requirements.

\begin{table}[!ht]
    \centering
    \caption{Space and time complexity for various methods used to calculate the audio sound field at a total of $N_x N_y N_z$ points generated by a MCPAL system.
    The field is sampled using $N_x,N_y,$ and $N_z$ mesh grid points in the $x$-, $y$-, and $z$-directions, respectively.
    Here, $N_\mathrm{G}$ is the order of the GBE, 
    $N_\mathrm{E}$ is the number of elements and $N_\mathrm{S}$ is the order of the SWE.}
    \vspace{1mm}
    \begin{tabular}{cccc}
        \toprule
        \textbf{Method} & \textbf{Total space complexity} & \textbf{Time complexity} \\
        \midrule
        DIM\cite{Zhong2021} & $\mathcal{O}(N_x N_y N_z)$ & $\mathcal{O}(M_x M_y N_x^2 N_y^2 N_z^2)$ \\
        Paraxial and non-paraxial GBE\cite{Zhuang2023, Zhu2023} & $\mathcal{O}(N_x N_y N_z)$ & $\mathcal{O}(N_\mathrm{E}^2 N_\mathrm{G}^2 N_x N_y N_z^2)$ \\
        SWE\cite{Zhuang2025} & $\mathcal{O}(N_x N_y N_z)$ & $\mathcal{O}(N_\mathrm{E}^2 N_\mathrm{S}^4 N_x N_y N_z^3)$ \\
        $k$-space (this work) & $\mathcal{O}(N_x N_y N_z)$ & $\mathcal{O}(N_x N_y N_z \log(N_x N_y N_z))$ \\
        \bottomrule
    \end{tabular}
    \label{tab:DiffSimu_complex}
\end{table}
It can be observed that the DIM has the highest time complexity. 
Moreover, while the GBE method employs simplifications for single field point calculations [$\mathcal{O}(N_\mathrm{E}^2 N_\mathrm{G}^2 N_z)$], its computational complexity becomes inferior to the proposed method when evaluating the full spatial field at $N_x N_y N_z$ grid points.
The SWE method can accurately calculate the audio sound across the entire space, but it is limited to computing the audio sound radiated by PAL arrays composed of circular elements. 
Additionally, the computational complexity of the SWE method is related to the number of elements $N_\mathrm{E}$; when the PAL array has a large number of elements $N_\mathrm{E}$, this method requires substantial computational resources. 
In contrast, the proposed $k$-space method's computational complexity is independent of both the number and shape of the elements. 
It requires the least computational resources among all methods for calculating the audio sound field distribution across the entire space. 

\section{Numerical results\label{sec:Numerical}}

Table\,\ref{tab:table1} lists the parameters used in the simulations, where the center frequency of the ultrasound is defined as ${f}_\mathrm{c}=\qty(f_1+f_2)/2$. 
The integration domain of the Rayleigh integral shown in Eq.\,(\ref{eq:ultra_rayleigh}) is set as $-3\text{ m} \leq x_\text{s}, y_\mathrm{s} \leq 3\text{ m}$.
Meanwhile, the integration domain of the volume integral shown in Eq.\,(\ref{eq:audio_volume}) is set to cover the major part of the ultrasonic beam, which is $-3\text{ m} \leq x_\text{v},y_\mathrm{v} \leq 3\text{ m}$ and $-8\text{ m} \leq z_\text{v} \leq 8\text{ m}$.
 
 \begin{table}[ht]
 \renewcommand\arraystretch{0.8}
 \caption{\label{tab:table1}Parameters used in the simulations.}
 \setlength{\tabcolsep}{14mm}
 \vskip3pt
 \begin{tabular}{cc}
 \bottomrule\bottomrule
  Parameters&Value \\  
  \bottomrule
  Ambient temperature & $T=20^\circ$C \\ 
  Relative humidity of air & $h_\text{r}=60\%$ \\ 
  Center frequency of the ultrasound & $f_\text{c}=40$\,kHz \\
  Attenuation coefficients of the ultrasound & $\alpha_\text{c}=0.15$ Np/m \\
  Amplitude of the velocity profile & $v_0=0.1$ m/s \\
  Nonlinearity coefficient & $\beta=1.2$ \\ 
 \bottomrule\bottomrule
 \end{tabular}
 \end{table}

Section\,\ref{sec:sound_2} evaluates the convergence and accuracy of the proposed method through simulations of sound fields generated by a circular PAL.
Section\,\ref{sec:sound_array} presents numerical results for an MCPAL system without or with the beam steering.
Finally, Sec.\,\ref{sec:Simulation_computation_efficiency} discusses the computational resource demands associated with the numerical simulation using the proposed method and conventional methods.

\subsection{Sound fields generated by a circular PAL}
\label{sec:sound_2}

To verify the convergence and accuracy of the proposed method, a circular PAL with a radius of $a=50\,\mathrm{mm}$ is considered in this section.
In this section, the audio sound computed by the DIM serves as the reference solution for the comparative analysis.

\subsubsection{Convergence analysis}

\begin{figure}[htb]
\includegraphics[width = 0.8\textwidth]{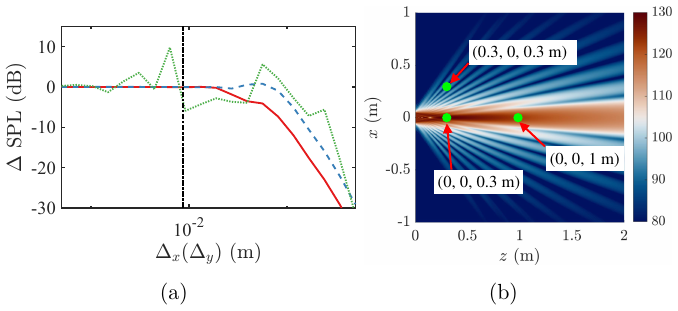}
\caption{
The numerical accuracy analysis of ultrasound computations using the ASA: 
(a) the predicted error in ultrasound pressure determined through the ASA across different mesh sizes ($\Delta_x(\Delta_y)$), the specified coordinates of the field points are: \SolidLine{color=BrewerSetRed}, $(0, 0, 0.3\text{ m})$; \SolidLine{color=BrewerSetBlue, dashed}, $(0, 0, 1\text{ m})$; \SolidLine{color=BrewerSetGreen, dotted}, $(0.3, 0, 0.3\text{ m})$; (b) the ultrasound calculated by the ASA with the mesh size $\Delta_x = \Delta_y = 2$\,mm. 
\DashdottedLine{}{27pt}, the wavelength (8.6\,mm) at 40\,kHz.
}
\label{fig:truncUltra}
\end{figure}

\begin{figure}[htb]
\includegraphics[width = 0.8\textwidth]{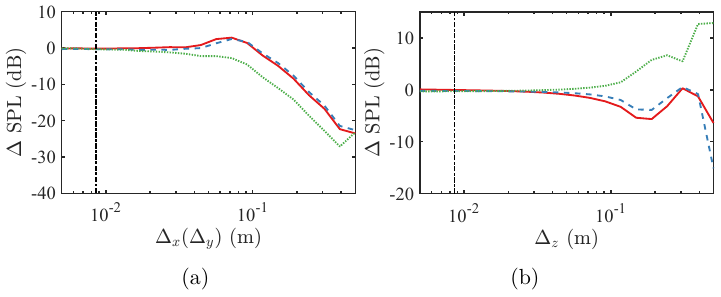}
\caption{
The predicted error in audio sound pressure determined through the proposed method across different mesh sizes includes: (a) $\Delta_x(\Delta_y)$; (b) $\Delta_z$.
The specified coordinates of the field points are: \SolidLine{color=BrewerSetRed}, $(0, 0, 0.3\text{ m})$; \SolidLine{color=BrewerSetBlue, dashed}, $(0, 0, 1\text{ m})$; \SolidLine{color=BrewerSetGreen, dotted}, $(0.3, 0, 0.3\text{ m})$.
\DashdottedLine{}{27pt}, the wavelength (8.6\,mm) at 40\,kHz.
}
\label{fig:trunc1}
\end{figure}

Figure\,\ref{fig:truncUltra} presents a numerical accuracy analysis of ultrasound at $f_\mathrm{c} = 40$\,kHz generated by a circular PAL computations using the ASA with different mesh sizes of $\Delta_x(\Delta_y)$ and the ultrasound calculated by the ASA with mesh sizes of $\Delta_x = \Delta_y = 2$\,mm.
It can be seen that the convergence of ASA depends on location relative to the main ultrasound beam. 
Within the primary beam region, solutions converge when the mesh size is smaller than $\lambda_\mathrm{c} \approx 8.6$\,mm (where $\lambda_\mathrm{c}$ is the wavelength of the center frequency of the ultrasound). 
However, in weaker peripheral regions where ultrasounds drop substantially, higher grid resolution (less than $\lambda_\mathrm{c}/2$) is required for convergence due to lower ultrasound pressure. 
As discussed in Sec.\,\ref{sec:theo_audio_sound}, the audio sound generated by a PAL system can be modeled as originating from a virtual volume source whose density is proportional to the product of two ultrasound fields [Eq.\,(\ref{eq:audio_volume})]. 
Since this represents a cumulative generation process, the accuracy requirements for ultrasound calculations are somewhat relaxed when computing the audio sound. 
Specifically, audio sound can be obtained accurately when the ultrasound field within the primary energy region is calculated with high accuracy, ensuring that the error in areas outside the main energy region remains under 10\,dB.
This relaxed accuracy threshold has been validated through convergence analysis of the audio field calculated by the proposed method, as demonstrated in Fig.\,\ref{fig:trunc1} and the following discussions.

Figure\,\ref{fig:trunc1} illustrates the predicted audio sound pressure error of the proposed method at several typical field points for a circular PAL with different mesh sizes of $\Delta_x(\Delta_y)$ when $\Delta_z = 10$\,mm and with different $\Delta_z$ when $\Delta_x(\Delta_y) = 10$\,mm. 
It can be observed that all error curves exhibit asymptotic convergence as the mesh size decreases, with $\Delta \mathrm{SPL}$ decreasing below 0.1\,dB when the mesh size is less than $\lambda_\mathrm{c} \approx 8.6 \text{ } \mathrm{mm}$. 
The results demonstrate that while the ASA requires strict grid resolution (less than $\lambda_\mathrm{c}/2$) for precise full-space ultrasound field calculations, the proposed method achieves accurate audio sound field calculations with relaxed spatial sampling (less than $\lambda_\mathrm{c}$).
The convergence behavior observed is consistent with the previously discussed results regarding the convergence of the primary energy region of ultrasound (Fig.\,\ref{fig:truncUltra}). 
These results confirm convergence and accuracy of the proposed method.
Given ultrasound's strong directivity, higher grid resolution is required along the propagation direction. 
To ensure simulation accuracy, $\Delta_x$ and $\Delta_y$ are set to 10\,mm, while $\Delta_z$ is set to 5\,mm for all following calculations.

\subsubsection{Audio sound field}

\begin{figure}[htb]
\includegraphics[width = 0.8\textwidth]{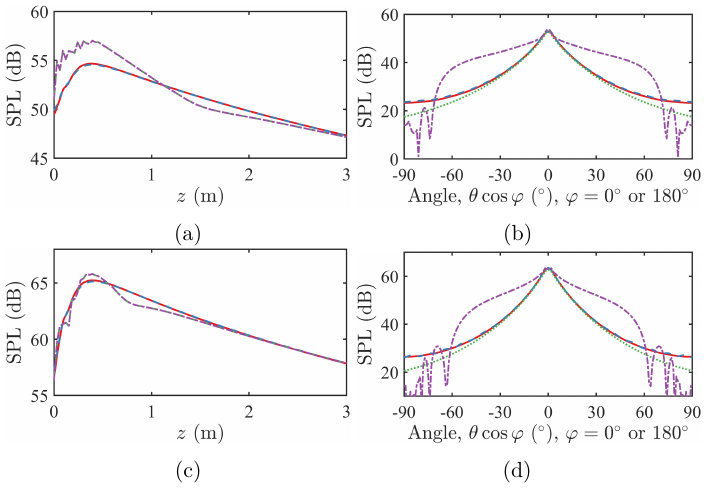}
\caption{
The audio sound pressure generated by a circular PAL. 
Left column: axial audio SPL; right column: angular audio SPL 1\,m away from $O$ on the $Oxz$ plane, 
The audio frequency is (top) 1\,kHz and (bottom) 2\,kHz. 
The audio sound pressure is obtained by: \SolidLine{color=BrewerSetRed}, the proposed method; \SolidLine{color=BrewerSetBlue, dashed}, the DIM method; \SolidLine{color=BrewerSetPurple, dash dot}, the paraxial GBE method; \SolidLine{color=BrewerSetGreen, dotted}, the non-paraxial GBE method. 
} 
\label{fig:audio1}
\end{figure}

Figure\,\ref{fig:audio1} illustrates the axial and angular audio sound generated by a circular PAL calculated by the proposed method, the DIM method, the paraxial GBE method and the non-paraxial GBE method, where $\theta$ denotes the zenithal angle ($0 \leq \theta \leq \pi$), measured from the positive $z$-axis and $\varphi$ represents the azimuthal angle ($0 \leq \varphi \leq 2 \pi$), 
measured in the $Oxy$ plane from the positive $x$-axis.
It can be observed that the axial audio sound pressure in the near field calculated by the paraxial and non-paraxial GBE method has an significant error compared to that calculated by the proposed method and the DIM method. 
This is because the paraxial approximation and non-paraxial approximation used in the GBE method result in significant errors at lower audio frequencies and in the near field.\cite{Cervenka2013, Zhuang2023}
Meanwhile, the audio sound pressure calculated by the proposed method has an error of less than 0.1\,dB compared to that calculated by the DIM method.
This is because the proposed method does not employ additional approximations when calculating the audio sound pressure; it merely utilizes the FFT to reduce the computational resources required for obtaining the quasilinear solution.
This simulation result validates the accuracy of the proposed method.

\subsection{Sound fields generated by a MCPAL system}
\label{sec:sound_array}

As shown in Fig.\,\ref{fig:sketch_PalArray}, two MCPAL systems with different arrangements is considered in this section. 
Both MCPAL systems consist of $24 \times 24$ circular PAL elements. 
The radius of the circular PAL element is 5\,mm.
The phased array technique enables the sound beam to be steered to the desired direction.
The unit vector denoting the beam direction is represented as $\vb{r}_\mathrm{d} = (\sin \theta_\mathrm{d} \cos \varphi_\mathrm{d}, \sin \theta_\mathrm{d} \sin \varphi_\mathrm{d}, \cos \theta_\mathrm{d})$, as illustrated in Figs.\,\ref{fig:sketch_PalArray}\,(b)\,and\,(c).
Then the complex weight of the $n$-th PAL element located at $\vb{r}_n = (x_n, y_n,0)$ of frequency $f_i$ can be expressed as $w_{i, n} = \exp(- \mathrm{i} k_i \vb{r}_n \vdot \vb{r}_\mathrm{d})$.


\subsubsection{Audio sound generated by a MCPAL system without the beam steering}
\label{sec:array_withoutSteer}
\begin{figure}[htb]
\includegraphics[width = 0.8\textwidth]{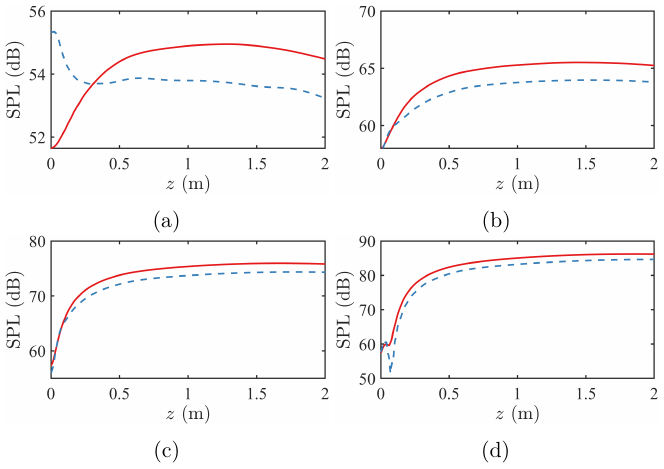}
\caption{
The axial audio sound pressure generated by the MCPAL system without the beam steering: 
\SolidLine{color=BrewerSetRed}, closely packed MCPAL system [Fig.~\ref{fig:sketch_PalArray}\,(b)]; \SolidLine{color=BrewerSetBlue,dashed}, uniformly spaced MCPAL system  [Fig.~\ref{fig:sketch_PalArray}\,(c)]. 
The audio frequency is (a) 500\,Hz, (b) 1\,kHz, (c) 2\,kHz, (d) 4\,kHz.
}
\label{fig:array1}
\end{figure}
\begin{figure}[htb]
\includegraphics[width = 0.8\textwidth]{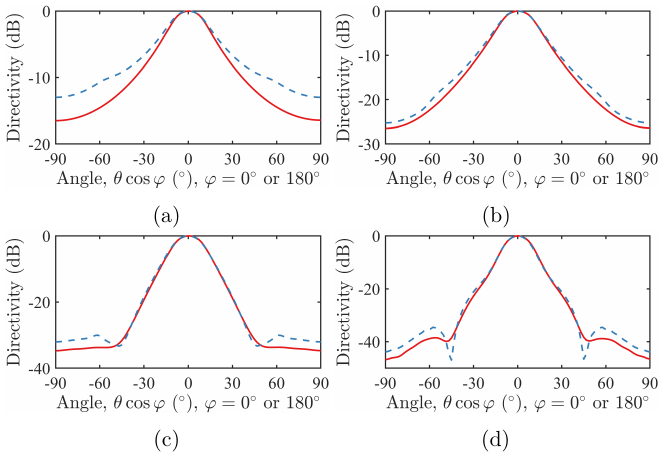}
\caption{
The angular normalized audio sound generated by the MCPAL system without the beam steering located at 0.5\,m away from $O$ on the $Oxz$ plane. 
\SolidLine{color=BrewerSetRed}, closely packed MCPAL system; \SolidLine{color=BrewerSetBlue,dashed}, uniformly spaced MCPAL system. 
The audio frequency is (a) 500\,Hz, (b) 1\,kHz, (c) 2\,kHz, (d) 4\,kHz.
}
\label{fig:array2}
\end{figure}
Figures\,\ref{fig:array1}\,and\,\ref{fig:array2} illustrate the simulated audio sound fields using the proposed method under the condition that all elements in the MCPAL systems have the same vibration velocity. 
Specifically, Fig.\,\ref{fig:array1} shows the axial audio sound pressure radiated by MCPAL systems with different arrangements, while Fig.\,\ref{fig:array2} displays the normalized audio sound pressure at different angles when positioned 0.5\,m away from the MCPAL systems. 
It can be observed that the axial audio sound pressure generated by the uniformly spaced MCPAL system is generally lower than that of the closely packed MCPAL system beyond certain distances. 
For example, at an audio frequency of 1\,kHz, the audio SPL generated by the uniformly spaced MCPAL system at 2\,m is 1.5\,dB lower than that of the closely packed MCPAL system at the same distance. 
The uniformly spaced MCPAL system demonstrates more pronounced interference effects than the closely packed MCPAL system, which results in reduced directivity.\cite{Zhong2023b} 
This conclusion is also corroborated by Fig.\,\ref{fig:array2}, which shows that the uniformly spaced MCPAL system exhibits a broader directivity pattern, leading to greater divergence of sound energy and consequently a reduction in axial sound pressure.
The results demonstrate that the closely packed MCPAL system not only achieves a more compact form factor (24.5\,cm\,$\times$\,20.9 cm versus 24 \,cm\,$\times$\,24\,cm for uniformly spaced MCPAL system), but also generates audio sound with both higher acoustic energy and superior directivity under identical operating conditions. 
The proposed method successfully reveals significant performance variations between different array configurations, highlighting the critical importance of element arrangement in MCPAL system design.

\subsubsection{Audio sound generated by a steerable MCPAL system}
\label{sec:array_steerable}

\begin{figure}[htb]
\includegraphics[width = 0.9\textwidth]{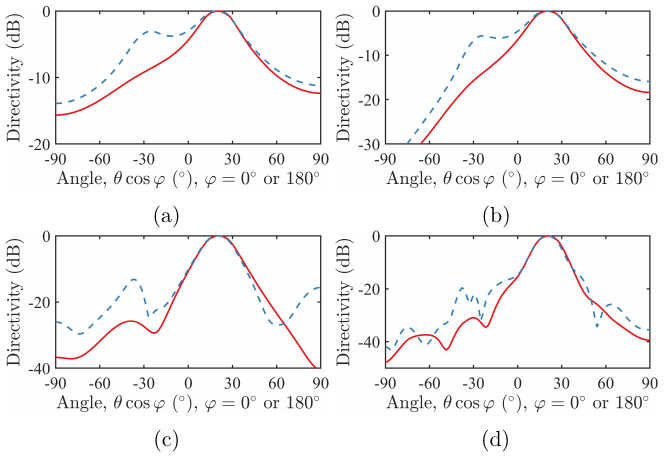}
\caption{
The angular normalized audio sound generated by the steerable MCPAL system located at 0.5\,m away from $O$ on the $Oxz$ plane. 
The steering angle is $20^\circ$ on the $Oxz$ plane.
\SolidLine{color=BrewerSetRed}, closely packed MCPAL system; \SolidLine{color=BrewerSetBlue,dashed}, uniformly spaced MCPAL system. 
The audio frequency is (a) 500\,Hz, (b) 1\,kHz, (c) 2\,kHz, (d) 4\,kHz.}
\label{fig:steer_array1}
\end{figure}

\begin{figure}[htb]
\includegraphics[width = 0.7\textwidth]{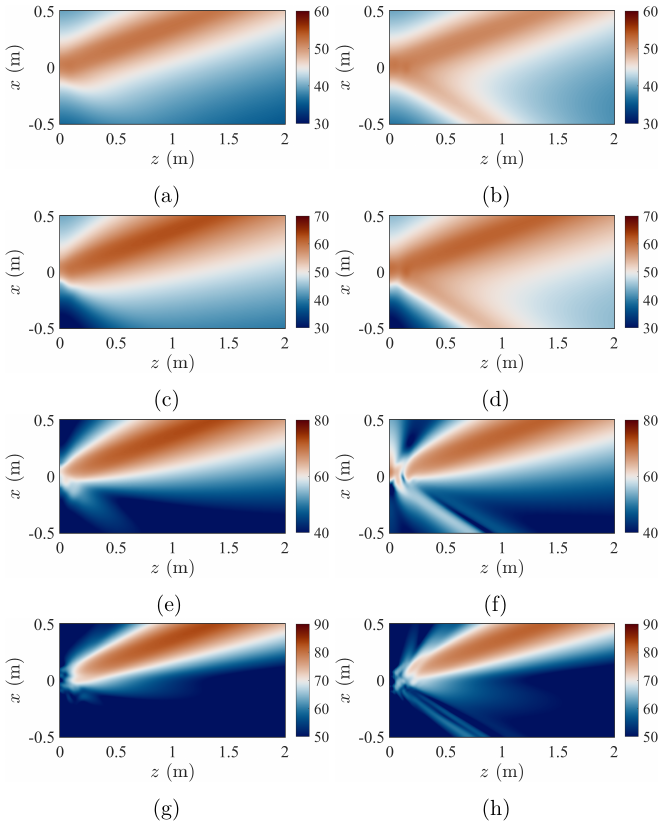}
\caption{
The audio sound field generated by a steerable MCPAL system on the $Oxz$ plane with steering angle $20^\circ$. 
Left column: closely packed MCPAL system; right column: uniformly spaced MCPAL system, 
the audio frequency is: (a--b) 500\,Hz; (c--d) 1\,kHz; (e--f) 2\,kHz; (g--h) 4\,kHz.}
\label{fig:steer_array2}
\end{figure}

Figures\,\ref{fig:steer_array1}\,and\,\,\ref{fig:steer_array2} illustrate the audio sound generated by a steerable MCPAL system at a steering angle of $20^\circ$ on the $Oxz$ plane, simulated using the proposed method. 
Specifically, Fig.\,\ref{fig:steer_array1} shows the normalized audio sound pressure at different angles when positioned 0.5\,m away from the steerable MCPAL systems with different arrangements, while Fig.\,\ref{fig:steer_array2} displays the audio sound pressure distribution in the $Oxz$ plane for the MCPAL systems with different arrangements. 
It can be observed that the uniformly spaced steerable MCPAL system exhibits higher side lobes compared to the closely packed steerable MCPAL system. 
For example, at an audio frequency of 2\,kHz, the side lobe of the uniformly spaced MCPAL system is located at $-36^\circ$ with a normalized SPL of $-13.2$\,dB, while the side lobe of the closely packed MCPAL system is located at $-39.6^\circ$ with a normalized SPL of $-25.8$\,dB. 
This is because the uniformly spaced MCPAL system exhibits stronger interference effects compared to the closely packed MCPAL system.\cite{Zhong2023b}
Additionally, it can be observed that at low audio frequencies ($f_\mathrm{a} \leq 2$\,kHz), the uniformly spaced MCPAL system produces audio sound with more pronounced side lobes compared to higher frequencies ($f_\mathrm{a} > 2$\,kHz).
This occurs because, at lower audio frequencies, the ultrasonic frequencies $f_1$ and $f_2$ are closer, resulting in similar wavelengths. 
As a result, the side lobes associated with the wavelengths of the two ultrasonic waves exhibit nearly equivalent angles and intensities, resulting in an increased density of virtual sources in the directions of the side lobes, and consequently, more pronounced audio side lobes.\cite{Shi2011, Zhong2023b}
In contrast, at higher audio frequencies, the wavelengths of the two ultrasonic waves differ significantly, causing their side lobes to vary in angle and intensity. 
This reduces the virtual source density in the side lobe directions, weakening the side lobe effect in the audio output compared to low-frequency conditions.

From the above simulation results, it can be observed that the closely packed MCPAL system exhibits smaller side lobes compared to the uniformly spaced MCPAL system, which is highly significant for precise sound field control. 
Moreover, the proposed method's capability to accurately simulate the audio sound of MCPAL systems enables effective differentiation of acoustic radiation patterns generated by different array configurations. 
This computational advantage of the proposed method offers significant benefits for the design and practical implementation of MCPAL systems.

\subsection{Computation efficiency}
\label{sec:Simulation_computation_efficiency}

As discussed in Sec.\,\ref{sec:complex_analy}, only the proposed method and the DIM possess the capability to simulate audio sound fields generated by arbitrary planar MCPAL systems. 
Consequently, this section presents a comparative analysis of the computational resource requirements between these two approaches when modeling the steerable MCPAL system mentioned in Sec.\,\ref{sec:array_steerable}.
Table\,\ref{tab:table2} shows the calculation time and the required memory of the proposed method and the DIM for the area $-3\text{ m} \leq x \leq 3\text{ m}$, $-3\text{ m} \leq y \leq 3\text{ m}$, and $-8\text{ m} \leq z \leq 8\text{ m}$ with $N_x N_y N_z = 601 \times 601 \times 3200 \approx 10^9$ points of the steerable MCPAL system. 
Due to the large number of field points in the full spatial domain (around $10^9$), the reported computation times for DIM represent single-point calculations scaled by the total field count ($N_x N_y N_z$).
The results are obtained on a computer with a 2.1\,GHz CPU and 1\,TB random access memory. 

\begin{table}[ht]
\renewcommand\arraystretch{0.8}
\caption{\label{tab:table2} Computational cost comparison for full-space audio sound calculation methods (steerable MCPAL system). The calculation time for the DIM is estimated by multiplying the single-point calculation time (value in parentheses) by the total number of field points ($N_x N_y N_z$).}
\setlength{\tabcolsep}{5mm}
\vskip3pt
\begin{tabular}{ccc}
\bottomrule\bottomrule
& Calculation time (s) & Required memory (GB) \\ 
\bottomrule
Proposed method & 317.7 & 34.5 \\
DIM & $(1.6 \times 10^5) \times N_xN_yN_z$ & 34.4 \\ 

\bottomrule\bottomrule
\end{tabular}
\end{table}

It can be observed that even for a single field point calculation, the DIM requires approximately $1.6 \times 10^5$ seconds of computation time due to its inherent need to evaluate five-fold integral.
In contrast, the proposed $k$-space method calculates the audio sound across the entire space generated by the MCPAL system in just $317.7$\,s.
Therefore, the proposed method holds a significant advantage when computing the audio sound radiated by MCPAL systems.

\section{Conclusion}
A $k$-space method for calculating audio sound generated by a MCPAL system is proposed in this article.
The method calculates the quasilinear solution of the Westervelt equation in a two-phase process.
First, the ultrasound field is computed employing the ASA technique. 
Subsequently, a frequency-domain $k$-space method is applied to diminish the computational expense associated with the volume integral necessary for audio sound computation. 
The $k$-space method, by leveraging the FFT, significantly curtails the computational load of the volume integral within the quasilinear solution framework.

The proposed method does not constrain the PAL's shape or the distribution of surface velocity. 
Consequently, this method is versatile, enabling the simulation of planar baffled PALs with diverse shapes and velocity distributions, including complex MCPAL systems composed of multiple elements, while maintaining low computational resource requirement. 
Furthermore, numerical results validate that the proposed $k$-space method exhibits superior accuracy compared to alternative methods and demonstrates enhanced computational efficiency when applied to MCPAL systems.
Additionally, simulation outcomes illustrate the method's capability to swiftly predict variations in the audio sound fields radiated by MCPAL systems with different element distributions. 
Accordingly, it is concluded that the proposed method outperforms the existing methods and provides a useful tool to analyze the audio sound field generated by a MCPAL system.

 \begin{acknowledgments}
T. Z. and J. L. gratefully acknowledge the
financial support by the National Natural Science Foundation of China (Grant No.~12274221), the AI \& AI for Science Project of Nanjing University, and the Special Project of the 2024 Changshu (Suzhou Acoustic Valley Innovation Fund), Special Project (Grant No. Changkehe [2024] 13).
 \end{acknowledgments}

\section*{AUTHOR DECLARATIONS}
\subsection*{Conflict of Interest}
The authors have no conflicts of interest to disclose.
\section*{DATA AVAILABILITY}
Additional data supporting the findings of this study are available from the corresponding author upon reasonable request.
 
 
 
 
 
 
 
 \clearpage
 \bibliographystyle{jasanum2.bst}
 \bibliography{Reference.bib}
 
 

 
 \end{document}